\documentclass[12pt]{article}

\usepackage{amsthm}
\usepackage{amsmath,amsfonts}
\usepackage{natbib}
\usepackage{graphicx}        
\usepackage{float}
\usepackage{subeqnarray}

\usepackage{url}
\usepackage{subcaption} 

\title{\Large Bayesian Computation with Intractable Likelihoods}
\author{Matthew T. Moores\thanks{National Institute for Applied Statistics Research Australia, School of Mathematics \& Applied Statistics, University of Wollongong, NSW 2522, Australia.} \and Anthony N. Pettitt\thanks{School of Mathematical Sciences,  Queensland University of Technology, Brisbane, Queensland 4001, Australia.} \and Kerrie Mengersen\footnotemark[2]}

\begin{document}
\maketitle

\begin{abstract}
This article surveys computational methods for posterior inference with intractable likelihoods, that is where the likelihood function is unavailable in closed form, or where evaluation of the likelihood is infeasible. We review recent developments in pseudo-marginal methods, approximate Bayesian computation (ABC), the exchange algorithm, thermodynamic integration, and composite likelihood, paying particular attention to advancements in scalability for large datasets. We also mention \textsf{R} and \textsf{MATLAB} source code for implementations of these algorithms, where they are available.
\end{abstract}

The likelihood function plays an important role in Bayesian inference, since it connects the observed data with the statistical model. Both simulation-based (e.g. MCMC) and optimisation-based (e.g. variational Bayes) algorithms require the likelihood to be evaluated pointwise, up to an unknown normalising constant. However, there are some situations where this evaluation is analytically and computationally intractable. For example, when the complexity of the likelihood grows at a combinatorial rate in terms of the number of observations, then likelihood-based inference quickly becomes infeasible for the scale of data that is regularly encountered in applications. 

Intractable likelihoods arise in a variety of contexts, including models for DNA mutation in population genetics \citep{Pritchard1999,Marjoram2003}, 
models for the spread of disease in epidemiology \citep{ONeill2000,McKinley2018}, models for the formation of galaxies in astronomy \citep{Cameron2012}, and estimation of the model evidence in Bayesian model choice \citep{Friel2008}. This chapter will mainly focus on Markov random field (MRF) models with discrete state spaces, such as the Ising, Potts, and exponential random graph models (ERGM). These models are used for image segmentation or analysis of social network data, two areas where millions of observations are commonplace. There is therefore a need for scalable inference algorithms that can handle these large volumes of data.

The Ising, Potts, or ERGM likelihood functions can be expressed in the form of an exponential family:
\begin{equation}\label{eq:like}
p(\vec{y} \mid \boldsymbol\theta) = \frac{\exp\left\{\boldsymbol\theta^T \vec{s}(\vec{y}) \right\}}{\mathcal{C}(\boldsymbol\theta)},
\end{equation}
where the observed data $\vec{y} = y_1, \dots, y_n$ is in the form of an undirected graph, $\boldsymbol\theta$ is a vector of unknown parameters, $\vec{s}(\vec{y})$ is a corresponding vector of jointly-sufficient statistics for these parameters, and $\mathcal{C}(\boldsymbol\theta)$ is an intractable normalising constant, also known as a partition function:
\begin{equation}\label{eq:partition}
\mathcal{C}(\boldsymbol\theta) = \sum_{\vec{y} \in \mathcal{Y}} \exp\left\{\boldsymbol\theta^T \vec{s}(\vec{y}) \right\},
\end{equation}
where the sum is over all possible configurations of states, $\vec{y} \in \mathcal{Y}$.

In the case of an Ising model, a single node can take one of two possible values, $y_i \in \{0, 1\}$. For example, in image analysis the value $1$ might represent a foreground pixel, while $0$ represents the background. The $q$-state Potts model generalises this construction to more than two states, so $y_i \in \{1, \dots, q\}$. The cardinality of the configuration space, $\#\mathcal{Y}$, is then $q^n$. Even with only 2 states and $n=100$ pixels, computation of \eqref{eq:partition} requires more than $10^{30}$ floating point operations. It would take a supercomputer with 100 PetaFLOPS over 400,000 years to find an answer.

Both the Ising and Potts models possess a single parameter, $\beta$, known as the inverse temperature. The corresponding sufficient statistic is then:
\begin{equation}
s(\vec{y}) = \sum_{i \sim \ell \in \mathcal{E}} \delta(y_i, y_\ell),
\end{equation}
where $\mathcal{E}$ is the set of all unique pairs of neighbours $i \sim \ell$ in the observed graph, and $\delta(x,y)$ is the Kronecker delta function, which equals $1$ when $x = y$ and $0$ otherwise. We assume a first-order neighbourhood structure, so a given pixel $y_i$ would have up to 4 neighbours in a regular 2D lattice, or 6 neighbours in 3D. Pixels on the boundary of the image domain have less than 4 (or 6) neighbours, so $\#\mathcal{E} = 2(n - \sqrt{n})$ for a square 2D lattice, or $3(n - n^{2/3})$ for a cube.

The observed data for an ERGM can be represented as a binary adjacency matrix $Y$, encoding the presence or absence of a neighbourhood relationship between nodes $i$ and $j$: $[Y]_{i,j} = 1$ if $i \sim j$; $[Y]_{i,j} = 0$ otherwise. $\#\mathcal{Y}$ for an ERGM is equal to $2^M$, where $M = n(n-1)/2$ is the maximum number of ties in an undirected graph with $n$ nodes. As with the Ising or Potts models, computing the normalising constant \eqref{eq:partition} is therefore intractable for non-trivial graphs. Various kinds of ERGM can be defined by the choice of sufficient statistics. The simplest example is the Bernoulli random graph \citep{ErdHos1959}, which has a single statistic $s_1(Y) = m$, the number of connected neighbours in the graph. In an undirected graph, this is half the number of nonzero entries in the adjacency matrix. An important class of graph statistics are the numbers of $k$-stars \citep{Frank1986}, which can be defined in terms of the degree distribution \citep{Olbrich2010}:
\begin{equation}
n_k = \sum_{i=1}^n \binom{d_i}{k},
\label{eqn:kstar}
\end{equation}
where the degree $d_i$ is the number of neighbours of node $i$:
\begin{equation}
d_i = \sum_{j=1}^n [Y]_{ij}.
\label{eqn:degree}
\end{equation}
Note that under this definition $n_1 = 2m$, since each tie is counted twice. An alternative definition, which avoids double-counting, is given by:
\begin{center}
\begin{tabular}{ll}
$n_1 = \sum_{i<j} [Y]_{ij}$ & number of edges\\
$n_2 = \sum_{i<j<k} [Y]_{ik} [Y]_{jk}$ & number of 2-stars\\
$n_3 = \sum_{i<j<l<k} [Y]_{ik} [Y]_{jk} [Y]_{lk}$ & number of 3-stars.
\end{tabular}
\end{center}


The remainder of this chapter will describe various MCMC methods that target the posterior distribution $\pi(\boldsymbol\theta \mid \vec{y})$, or some approximation thereof. This will be in the context of a random walk Metropolis (RWM) algorithm that proposes a new value of $\boldsymbol\theta'$ at iteration $t$ using a (multivariate) Gaussian proposal distribution, $q(\cdot \mid \boldsymbol\theta_{t-1}) \sim \mathcal{N}(\boldsymbol\theta_{t-1}, \Sigma_t)$. Methods for tuning the proposal bandwidth $\Sigma_t$ have been described by \citet{Andrieu2008} and \citet{Roberts2009}. Normally, the proposed parameter value would be accepted with probability $\min\{1, \rho_t\}$, or else rejected, where $\rho_t$ is the Radon--Nikod{\'y}m derivative:
\begin{equation}\label{eq:mhRatio}
\rho_t = \frac{q\left(\boldsymbol\theta^{(t-1)} \mid \boldsymbol\theta'\right) p\left(\vec{y} \mid \boldsymbol\theta'\right) \pi_0\left(\boldsymbol\theta'\right)}{q\left(\boldsymbol\theta' \mid \boldsymbol\theta^{(t-1)}\right) p\left(\vec{y} \mid \boldsymbol\theta^{(t-1)}\right) \pi_0\left(\boldsymbol\theta^{(t-1)}\right)},
\end{equation}
$\pi_0(\boldsymbol\theta)$ is the prior density for the parameter/s, and $p\left(\vec{y} \mid \boldsymbol\theta\right)$ is the likelihood \eqref{eq:like}. If we use a symmetric proposal distribution $q$ and a uniform prior $\pi_0$, then these terms will cancel, leaving:
\begin{equation}\label{eq:ratio}
\rho_t = \frac{\psi\left(\vec{y} \mid \boldsymbol\theta'\right)}{\psi(\vec{y} \mid \boldsymbol\theta^{(t-1)})} \frac{\mathcal{C}(\boldsymbol\theta^{(t-1)})}{\mathcal{C}\left(\boldsymbol\theta'\right)},
\end{equation}
which is the ratio of unnormalised likelihoods $\psi = \exp\left\{\boldsymbol\theta^T \vec{s}(\vec{y})\right\}$, multiplied by the ratio of intractable normalising constants \eqref{eq:partition}. It is clearly infeasible to evaluate \eqref{eq:ratio} directly, so alternative algorithms are required. One option is to estimate $\rho_t$ by simulation, which we categorise as auxiliary variable methods: pseudo-marginal algorithms, the exchange algorithm, and approximate Bayesian computation (ABC). Other methods include path sampling, also known as thermodynamic integration (TI), pseudolikelihood and composite likelihood. 

\section{Auxiliary Variable Methods}

\subsection{Pseudo-Marginal Algorithms}

Pseudo-marginal algorithms \citep{Beaumont2003,Andrieu2009} are computational methods for fitting latent variable models, that is where the observed data $\vec{y}$ can be considered as noisy observations of some unobserved or hidden states, $\vec{x}$. For example, hidden Markov models (HMMs) are commonly used in time series analysis and signal processing. Models of this form can also arise as the result of data augmentation approaches, such as for mixture models \citep{Dempster1977,Tanner1987}. The marginal likelihood is of the following form:
\begin{equation}
p(\vec{y} \mid \boldsymbol\theta) = \int_\mathcal{X} p(\vec{y} \mid \vec{x}) \,p(\vec{x} \mid \boldsymbol\theta) \,d\vec{x},
\end{equation}
which can be intractable if the state space is very high-dimensional and non-Gaussian. In this case, we can substitute an unbiased, non-negative estimate of the likelihood. 

\citet{ONeill2000} introduced the Monte Carlo within Metropolis (MCWM) algorithm, which replaces both $p\left(\vec{y} \mid \boldsymbol\theta'\right)$ and $p\left(\vec{y} \mid \boldsymbol\theta^{(t-1)}\right)$ in the Metropolis-Hastings ratio $\rho_t$ \eqref{eq:mhRatio} with importance sampling estimates:
\begin{equation}\label{eq:is}
\tilde{p}_{IS}(\vec{y} \mid \boldsymbol\theta) \approx \frac{1}{M} \sum_{m=1}^M p(\vec{y} \mid X_m) \frac{p(X_m \mid \boldsymbol\theta)}{q(X_m \mid \boldsymbol\theta)},
\end{equation}
where the samples $X_1, \dots, X_M$ are drawn from a proposal distribution $q(X_m \mid \boldsymbol\theta)$ for $\boldsymbol\theta'$ and $\boldsymbol\theta^{(t-1)}$. MCWM is generally considered as an approximate algorithm, since it does not target the exact posterior distribution for $\boldsymbol\theta$. However, \citet{Medina-Aguayo2016} have established some conditions under which MCWM converges to the correct target distribution as $M \rightarrow \infty$. See also \citet{Nicholls2012} and \citet{Alquier2014} for further theoretical analysis of approximate pseudo-marginal methods.

\citet{Beaumont2003} introduced the grouped independence Metropolis-Hastings (GIMH) algorithm, which does target the exact posterior. The key difference is that $\tilde{p}_{IS}\left(\vec{y} \mid \boldsymbol\theta^{(t-1)}\right)$ is reused from the previous iteration, rather than being recalculated every time. The theoretical properties of this algorithm have been an active area of research, with notable contributions by \citet{Andrieu2009,Maire2014,Andrieu2015}, and \citet{Sherlock2015a}. \citet{Andrieu2010} introduced the particle MCMC algorithm, which is a pseudo-marginal method that uses sequential Monte Carlo (SMC) in place of importance sampling. This is particularly useful for HMMs, where SMC methods such as the bootstrap particle filter provide an unbiased estimate of the marginal likelihood \citep{Pitt2012}. Although importance sampling and SMC are both unbiased estimators, it is necessary to use a large enough value of $M$ so that the variance is kept at a reasonable level. Otherwise, the pseudo-marginal algorithm can fail to be variance-bounding or geometrically ergodic \citep{Lee2014}. \citet{Doucet2014} recommend choosing $M$ so that the standard deviation of the log-likelihood estimator is between 1 and 1.7.

Pseudo-marginal algorithms can be computationally-intensive, particularly for large values of $M$. One strategy to reduce this computational burden, known as the Russian Roulette algorithm \citep{Lyne2015}, is to replace $\tilde{p}_{IS}(\vec{y} \mid \boldsymbol\theta)$ \eqref{eq:is} with a truncated infinite series:
\begin{equation}\label{eq:roulette}
\tilde{p}_{RR}(\vec{y} \mid \boldsymbol\theta) = \sum_{j=0}^\tau V_{\boldsymbol\theta}^{(j)},
\end{equation}
where $\tau$ is a random stopping time and $V_{\boldsymbol\theta}^{(j)}$ are random variables such that \eqref{eq:roulette} is almost surely finite and $\mathbb{E}[\tilde{p}_{RR}(\vec{y} \mid \boldsymbol\theta)] = p(\vec{y} \mid \boldsymbol\theta)$. There is a difficulty with this method, however, in that the likelihood estimates are not guaranteed to be non-negative. \citet{Jacob2015} have established that there is no general solution to this sign problem, although successful strategies have been proposed for some specific models.

Another important class of algorithms for accelerating pseudo-marginal methods involve approximating the intractable likelihood function using a surrogate model. For example, the delayed-acceptance (DA) algorithm of \citet{Christen2005} first evaluates the Metropolis-Hastings ratio \eqref{eq:mhRatio} using a fast, approximate likelihood $\tilde{p}_{DA}(\vec{y} \mid \boldsymbol\theta)$. The proposal $\boldsymbol\theta'$ is rejected at this screening stage with probability $1 - \min\{1, \rho_t\}$. Otherwise, a second ratio $\rho^{(2)}_{DA}$ is calculated using a full evaluation of the likelihood function \eqref{eq:is}. The acceptance probability $\min\{1, \rho^{(2)}_{DA}\}$ is modified at the second stage according to:
\begin{equation}
\rho_{DA}^{(2)} = \frac{\tilde{p}_{IS}(\vec{y} \mid \boldsymbol\theta') \,\pi_0(\boldsymbol\theta')}{\tilde{p}_{IS}(\vec{y} \mid \boldsymbol\theta^{(t-1)}) \,\pi_0(\boldsymbol\theta^{(t-1)})} \frac{\tilde{p}_{DA}(\vec{y} \mid \boldsymbol\theta^{(t-1)}) \,\pi_0(\boldsymbol\theta^{(t-1)})}{\tilde{p}_{DA}(\vec{y} \mid \boldsymbol\theta') \,\pi_0(\boldsymbol\theta')},
\end{equation}
which corrects for the conditional dependence on acceptance at the first stage and therefore preserves the exact target distribution. DA has been used for PMCMC by \citet{Golightly2015}, where the linear noise approximation \citep{Fearnhead2014} was used for $\tilde{p}_{DA}$. \citet{Sherlock2015} instead used $k$-nearest-neighbours for $\tilde{p}_{DA}$ in a pseudo-marginal algorithm.

\citet{Drovandi2015} proposed an approximate pseudo-marginal algorithm, using a Gaussian process (GP) as a surrogate log-likelihood. The GP is trained using a pilot run of MCWM, then at each iteration $\log \tilde{p}(\vec{y} \mid \boldsymbol\theta')$ is either approximated using the GP or else using SMC or importance sampling, depending on the level of uncertainty in the surrogate model for $\boldsymbol\theta'$. \textsf{MATLAB} source code is available from \url{http://www.runmycode.org/companion/view/2663}. \citet{Stuart2016} have shown that, under certain assumptions, a GP provides a consistent estimator of the negative log-likelihood, and they provide error bounds on the approximation.

\subsection{Exchange Algorithm}

\citet{Moeller2006} introduced a MCMC algorithm for the Ising model that targets the exact posterior distribution for $\beta$. An auxiliary variable $\vec{x}$ is defined on the same state space as $\vec{y}$, so that $\vec{x}, \vec{y} \in \mathcal{Y}$. This is a data augmentation approach, where we simulate from the joint posterior $\pi(\beta, \vec{x} \mid \vec{y})$, which admits the posterior for $\beta$ as its marginal. Given a proposed parameter value $\beta'$, a proposal $\vec{x}'$ is simulated from the model to obtain an unbiased sample from \eqref{eq:like}. This requires perfect simulation methods, such as coupling from the past \citep{Propp1996}, perfect slice sampling \citep{Mira2001}, or bounding chains \citep{Huber2003,Butts2018}. Refer to \citet{Huber2016} for further explanation of perfect simulation. Instead of \eqref{eq:ratio}, the joint ratio for $\beta'$ and $\vec{x}'$ becomes:
\begin{equation}\label{eq:MAVM}
\rho_t = \frac{\psi\left(\vec{y} \mid \beta'\right)}{\psi\left(\vec{y} \mid \beta^{(t-1)}\right)} \frac{\psi\left(\vec{x}' \mid \tilde\beta\right)}{\psi\left(\vec{x}^{(t-1)} \mid \tilde\beta\right)} \frac{\psi(\vec{x}^{(t-1)} \mid \beta^{(t-1)})}{\psi\left(\vec{x}' \mid \beta'\right)},
\end{equation}
where the normalising constants $\mathcal{C}(\beta')$ and $\mathcal{C}(\beta^{(t-1)})$ cancel out with each other. This is analogous to an importance-sampling estimate of the  normalising constant with $M=1$ samples, since:
\begin{equation}
\mathbb{E}_{\vec{x}}\left[ \frac{\psi\left(\vec{x} \mid \beta\right)}{q(\vec{x} \mid \beta)} \right] = \mathcal{C}(\beta),
\end{equation}
where the proposal distribution $q(\vec{x} \mid \beta)$ is \eqref{eq:like}. This algorithm is therefore closely-related with pseudo-marginal methods such as GIMH.

 \citet{Murray2006} found that \eqref{eq:MAVM} could be simplified even further, removing the need for a fixed value of $\tilde\beta$. The exchange algorithm replaces \eqref{eq:ratio} with the ratio:
\begin{equation}\label{eq:exch}
\rho_t = \frac{\psi\left(\vec{y} \mid \beta'\right)}{\psi\left(\vec{y} \mid \beta^{(t-1)}\right)} \frac{\psi(\vec{x}' \mid \beta^{(t-1)})}{\psi\left(\vec{x}' \mid \beta'\right)}.
\end{equation}
However, perfect sampling is still required to simulate $\vec{x}'$ at each iteration, which can be infeasible when the state space is very large. \citet{Cucala2009} proposed an approximate exchange algorithm (AEA) by replacing the perfect sampling step with 500 iterations of Gibbs sampling. \citet{Caimo2011} were the first to employ AEA for fully-Bayesian inference on the parameters of an ERGM. AEA for the hidden Potts model is implemented in the \textsf{R} package `\texttt{bayesImageS}' \citep{bayesImageS} and AEA for ERGM is implemented in `\texttt{Bergm}' \citep{Caimo2014}.

\subsection{Approximate Bayesian Computation}

Like the exchange algorithm, ABC uses an auxiliary variable $\vec{x}$ to decide whether to accept or reject the proposed value of $\boldsymbol\theta'$. In the terminology of ABC, $\vec{x}$ is referred to as ``pseudo-data.'' Instead of a Metropolis-Hastings ratio such as \eqref{eq:ratio}, the summary statistics of the pseudo-data and the observed data are directly compared. The proposal is accepted if the distance between these summary statistics is within the ABC tolerance, $\epsilon$. This produces the following approximation:
\begin{equation}
\label{eq:abc_posterior}
p\left(\boldsymbol\theta \mid \vec{y} \right) \;\approx\; \pi_\epsilon\left(\boldsymbol\theta \mid \| \vec{s}(\vec{x}) - \vec{s}(\vec{y}) \| < \epsilon\right),
\end{equation}
where $\| \cdot \|$ is a suitable norm, such as Euclidean distance. Since $\vec{s}(\vec{y})$ are jointly-sufficient statistics for Ising, Potts, or ERGM, the ABC approximation (\ref{eq:abc_posterior}) approaches the true posterior as $n \to \infty$ and $\epsilon \to 0$. In practice there is a tradeoff between the number of parameter values that are accepted and the size of the ABC tolerance. 

\citet{Grelaud2009} were the first to use ABC to obtain an approximate posterior for $\beta$ in the Ising/Potts model. \citet{Everitt2012} used ABC within sequential Monte Carlo (ABC-SMC) for Ising and ERGM. ABC-SMC uses a sequence of target distributions $\pi_{\epsilon_t} \left(\boldsymbol\theta \mid \| \vec{s}(\vec{x}) - \vec{s}(\vec{y}) \| < \epsilon_t \right)$ such that $\epsilon_1 > \epsilon_2 > \dots > \epsilon_T$, where the number of SMC iterations $T$ can be determined dynamically using a stopping rule. The ABC-SMC algorithm of \citet{Drovandi2011} uses multiple MCMC steps for each SMC iteration, while the algorithm of \citet{DelMoral2012} uses multiple replicates of the summary statistics for each particle. \citet{Everitt2012} has provided a MATLAB implementation of ABC-SMC with the online supplementary material accompanying his paper.

The computational efficiency of ABC is dominated by the cost of drawing updates to the auxiliary variable, as reported by \citet{Everitt2012}. Thus, we would expect that the execution time for ABC would be similar to AEA or pseudo-marginal methods. Various approaches to improving this runtime have recently been proposed. ``Lazy ABC'' \citep{Prangle2014} involves early termination of the simulation step at a random stopping time, hence it bears some similarities with Russian Roulette. Surrogate models have also been applied in ABC, using a method known as Bayesian indirect likelihood \citep[BIL; ][]{Drovandi2011a,Drovandi2014}. Gaussian processes (GPs) have been used as surrogate models by \citet{Wilkinson2014} and \citet{Meeds2014}. \citet{Jaervenpaeae2016} used a heteroskedastic GP model and demonstrated how the output of the precomputation step could be used for Bayesian model choice. \citet{Moores2014} introduced a piecewise linear approximation for ABC-SMC with Ising/Potts models. \citet{Boland2017} derived a theoretical upper bound on the bias introduced by this and similar piecewise approximations. They also developed a piecewise linear approximation for ERGM. \citet{Moores2015a} introduced a parametric functional approximate Bayesian (PFAB) algorithm for the Potts model, which is a form of BIL where $\tilde{p}_{BIL}(\vec{y} \mid \boldsymbol\theta)$ is derived from an integral curve.

\section{Other Methods}

\subsection{Thermodynamic Integration}

\begin{figure}[H]
\includegraphics[width=\textwidth]{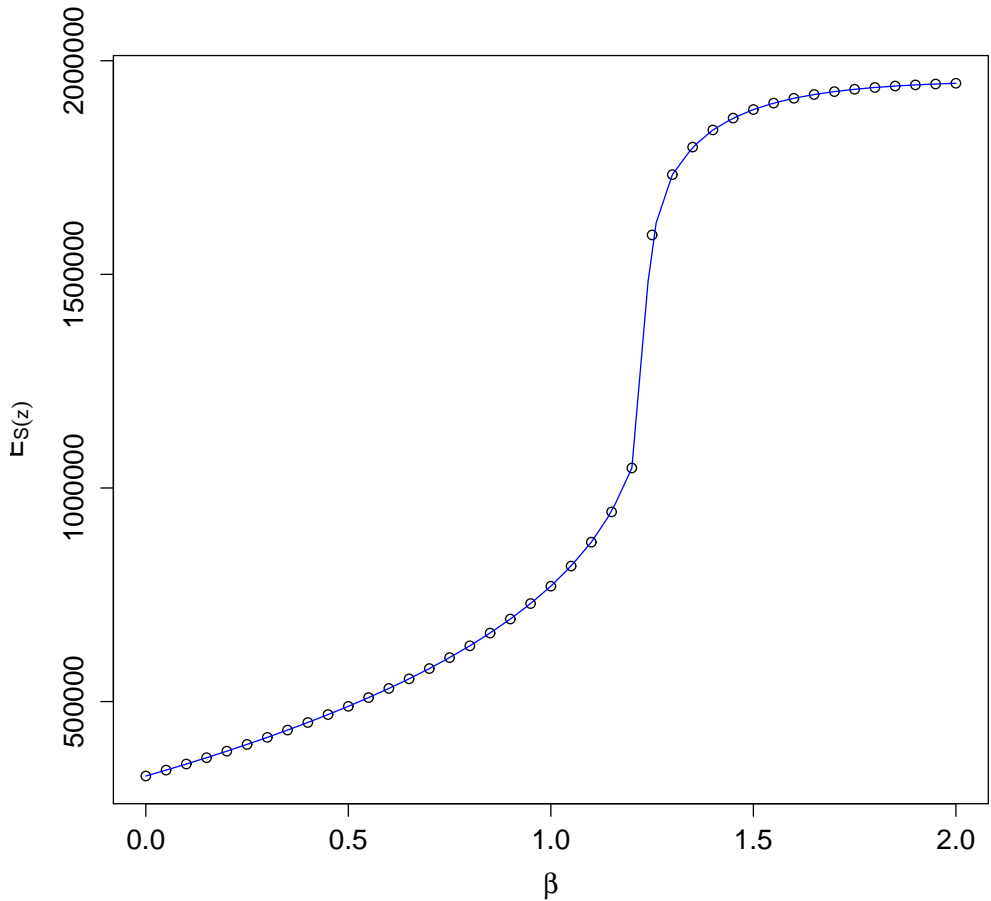}
\caption{Approximation of $\mathbb{E}_{\vec{y} | \beta}[s(\vec{y})]$ by simulation for fixed values of $\beta$, with linear interpolation.}
\label{f:path2d}       
\end{figure}

Since the Ising, Potts, and ERGM are all exponential families of distributions, the expectation of their sufficient statistic/s can be expressed in terms of the normalising constant:
\begin{equation}\label{eq:score}
\mathbb{E}_{\vec{y} | \boldsymbol\theta}[\vec{s}(\vec{y})] = \frac{\mathrm{d}}{\mathrm{d}\boldsymbol\theta} \log\{\mathcal{C}(\boldsymbol\theta)\}.
\end{equation}
\citet{Gelman1998} derived an approximation to the log-ratio of normalising constants for the Ising/Potts model, using the path sampling identity:
\begin{equation}
\label{eq:path}
\log\left\{\frac{\mathcal{C}(\beta_{t-1})}{\mathcal{C}(\beta')}\right\} = \int_{\beta'}^{\beta_{t-1}} \mathbb{E}_{\vec{y} | \beta}[s(\vec{y})] \, \mathrm{d}\beta ,
\end{equation}
which follows from \eqref{eq:score}.
The value of the expectation can be estimated by simulating from the Gibbs distribution \eqref{eq:like} for fixed values of $\beta$. At each iteration, $\log\{\rho_t\}$ \eqref{eq:ratio} can then be approximated by numerical integration methods, such as Gaussian quadrature or the trapezoidal rule. Figure~\ref{f:path2d} illustrates linear interpolation of $\mathbb{E}_{\vec{y} | \beta}[s(\vec{y})]$ on a 2D lattice for $q=6$ labels and $\beta$ ranging from 0 to 2 in increments of 0.05. This approximation was precomputed using the algorithm of \citet{Swendsen1987}. 

TI is explained in further detail by \citet[chap. 5]{Chen2000}. A reference implementation in \textsf{R} is available from the website accompanying \citet{Marin2007}. \citet{Friel2008} introduced the method of power posteriors to estimate the marginal likelihood or model evidence using TI. \citet{Calderhead2009} provide bounds on the discretisation error and derive an optimal temperature schedule by minimising the variance of the Monte Carlo estimate. \citet{Oates2016} introduced control variates for further reducing the variance of TI.

The TI algorithm has an advantage over auxiliary variable methods because the additional simulations are performed prior to fitting the model, rather than at each iteration. This is particularly the case when analysing multiple images that all have approximately the same dimensions. Since these simulations are independent, they can make use of massively parallel hardware. 
However, the computational cost is still slightly higher than pseudolikelihood, which does not require a pre-computation step.

\subsection{Composite Likelihood}

Pseudolikelihood is the simplest of the methods that we have considered and also the fastest. \citet{Ryden1998} showed that the intractable distribution \eqref{eq:like} could be approximated using the product of the conditional densities:
\begin{equation}
\label{eq:pseudo}
\tilde{p}_{PL}(\vec{y} \mid \boldsymbol\theta) \approx \prod_{i=1}^n p(y_i \mid y_{\setminus i}, \boldsymbol\theta).
\end{equation}
This enables the Metropolis-Hastings ratio $\rho_t$ \eqref{eq:mhRatio} to be evaluated using \eqref{eq:pseudo} to approximate both $p\left(\vec{y} \mid \boldsymbol\theta'\right)$ and $p\left(\vec{y} \mid \boldsymbol\theta^{(t-1)}\right)$ at each iteration. The conditional density function for the Ising/Potts model is given by:
\begin{equation}\label{eq:condPotts}
p(y_i \mid y_{\setminus i}, \beta) = \frac{\exp\left\{\beta\sum_{\ell \in \partial(i)}\delta(z_i,z_\ell)\right\}}{\sum_{j=1}^k \exp\left\{\beta\sum_{\ell \in \partial(i)}\delta(j,z_\ell)\right\}},
\end{equation}
where $\ell \in \partial(i)$ are the first-order (nearest) neighbours of pixel $i$. The conditional density for an ERGM is given by the logistic function:
\begin{equation}\label{eq:condERGM}
p([Y]_{ij} = 1 \mid [Y]_{\setminus ij}, \boldsymbol\theta) = \mathrm{logit}^{-1}\left\{\boldsymbol\theta^T \vec{s}(Y) \right\}.
\end{equation}
\begin{figure}
        \centering
        \begin{subfigure}{0.6\textwidth}
                \includegraphics[width=\textwidth]{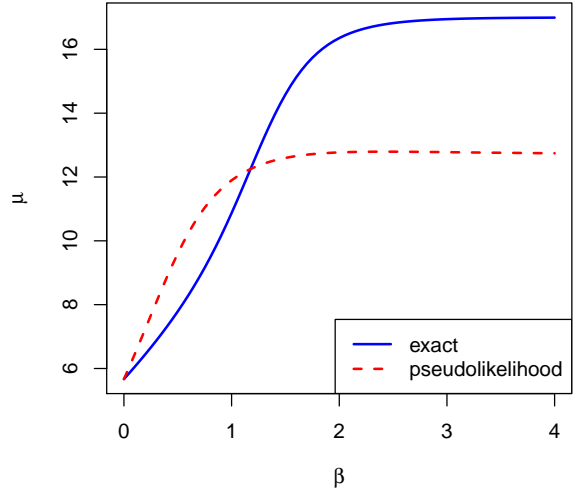}
                \caption{Expectation.}
                \label{f:ch6_pl_exp}
        \end{subfigure}%
\qquad
        \begin{subfigure}{0.6\textwidth}
                \includegraphics[width=\textwidth]{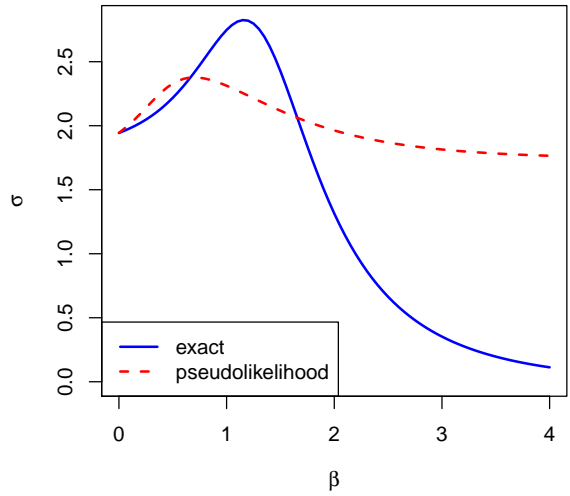}
                \caption{Standard deviation.}
                \label{f:ch6_pl_sd}
        \end{subfigure}%
\caption[Approximation error of pseudolikelihood.]{Approximation error of pseudolikelihood for $n=12,\,q=3$ in comparison to the exact likelihood calculated using a brute force method: (a) $\sum_{\vec{y} \in \mathcal{Y}} s(\vec{y}) p(\vec{y} | \beta)$ using either Equation~(\ref{eq:like}) or (\ref{eq:pseudo}); (b) $\sqrt{\sum_{\vec{y} \in \mathcal{Y}} \left( s(\vec{y}) - \mathbb{E}_{\vec{y}|\beta}[s(\vec{y})] \right)^2 p(\vec{y} | \beta)}$} 
\label{f:ch6_pl}
\end{figure}
Pseudolikelihood is exact when $\boldsymbol\theta=0$ and provides a reasonable approximation for small values of the parameters. However, the approximation error increases rapidly for the Potts/Ising model as $\beta$ approaches the critical temperature, $\beta_{crit}$, as illustrated by Figure~\ref{f:ch6_pl}. This is due to long-range dependence between the labels, which is inadequately modelled by the local approximation. Similar issues can arise for ERGM, which can also exhibit a phase transition.

\citet{Ryden1998} referred to Equation~\eqref{eq:pseudo} as point pseudolikelihood, since the conditional distributions are computed for each pixel individually. They suggested that the accuracy could be improved using block pseudolikelihood. This is where the likelihood is calculated exactly for small blocks of pixels, then \eqref{eq:pseudo} is modified to be the product of the blocks:
\begin{equation}
\tilde{p}_{BL}(\vec{y} \mid \boldsymbol\theta) \approx \prod_{i=1}^{N_B} p(\vec{y}_{B_i} | \vec{y}_{\setminus B_i}, \boldsymbol\theta)
\label{eq:ch6_pl_comp}
\end{equation}
where $N_B$ is the number of blocks, $\vec{y}_{B_i}$ are the labels of the pixels in block $B_i$, and $\vec{y}_{\setminus B_i}$ are all of the labels except for $\vec{y}_{B_i}$. This is a form of composite likelihood, where the likelihood function is approximated as a product of simplified factors \citep{Varin2011}. \citet{Friel2012} compared point pseudolikelihood to composite likelihood with blocks of $3 \times 3$, $4 \times 4$, $5 \times 5$, and $6 \times 6$ pixels. \citeauthor{Friel2012} showed that (\ref{eq:ch6_pl_comp}) outperformed (\ref{eq:pseudo}) for the Ising ($q=2$) model with $\beta < \beta_{crit}$. \citet{Okabayashi2011} discuss composite likelihood for the Potts model with $q > 2$ and have provided an open source implementation in the \textsf{R} package `\texttt{potts}' \citep{Geyer2014}.

Evaluating the conditional likelihood in (\ref{eq:ch6_pl_comp}) involves the normalising constant for $\vec{y}_{B_i}$, which is a sum over all of the possible configurations $\mathcal{Y}_{B_i}$. This is a limiting factor on the size of blocks that can be used. The brute force method that was used to compute Figure~\ref{f:ch6_pl} is too computationally intensive for this purpose. \citet{Pettitt2003} showed that the normalising constant can be calculated exactly for a cylindrical lattice by computing eigenvalues of a $k^r \times k^r$ matrix, where $r$ is the smaller of the number of rows or columns. The value of \eqref{eq:partition} for a free-boundary lattice can then be approximated using path sampling. \citet{Friel2004} extended this method to larger lattices using a composite likelihood approach.

The reduced dependence approximation (RDA) is another form of composite likelihood. \citet{Reeves2004} introduced a recursive algorithm to calculate the normalising constant using a lag-$r$ representation. \citet{Friel2009} divided the image lattice into sub-lattices of size $r_1 < r$, then approximated the normalising constant of the full lattice using RDA:
\begin{equation}
\mathcal{C}(\beta) \approx \frac{\mathcal{C}_{r_1 \times n}(\beta)^{r - r_1 + 1}}{\mathcal{C}_{r_1 - 1 \times n}(\beta)^{r - r_1}}
\label{eq:ch6_rda}
\end{equation}
\citet{McGrory2009} compared RDA to pseudolikelihood and the exact method of \citet{Moeller2006}, reporting similar computational cost to pseudolikelihood but with improved accuracy in estimating $\beta$. \citet{Ogden2017} showed that if $r$ is chosen proportional to $n$, then RDA gives asymptotically valid inference when $\beta < \beta_{crit}$. However, the error increases exponentially as $\beta$ approaches the phase transition. This is similar to the behaviour of pseudolikelihood in Figure~\ref{f:ch6_pl}. Source code for RDA is available in the online supplementary material for \citet{McGrory2012}.

\section{Conclusion}
This chapter has reviewed a variety of computational methods for Bayesian inference with intractable likelihoods. Auxiliary variable methods, such as the exchange algorithm and pseudo-marginal algorithms, target the exact posterior distribution. However, their computational cost can be prohibitive for large datasets. Algorithms such as delayed acceptance, Russian Roulette, and ``lazy ABC'' can accelerate inference by reducing the number of auxiliary variables that need to be simulated, without modifying the target distribution. Bayesian indirect likelihood (BIL) algorithms approximate the intractable likelihood using a surrogate model, such as a Gaussian process or piecewise function. As with thermodynamic integration, BIL can take advantage of a precomputation step to train the surrogate model in parallel. This enables these methods to be applied to much larger datasets by managing the tradeoff between approximation error and computational cost.

\section*{Acknowledgements}
This research was conducted by the Australian Research Council Centre of Excellence for Mathematical and Statistical Frontiers (project number CE140100049) and funded by the Australian Government.

\bibliographystyle{abbrvnat}
\bibliography{iLike}

\end{document}